\documentstyle[12pt]{article}

\catcode`\@=11
\long\def\@makefntext#1{ 
\protect\noindent \hbox to 3.2pt {\hskip-.9pt
$^{{\ninerm\@thefnmark}}$\hfil}#1\hfill} 
\def\thefootnote{\fnsymbol{footnote}}
 \def\@makefnmark{\hbox to 0pt{$^{\@thefnmark}$\hss}}  

\def\ps@myheadings{\let\@mkboth\@gobbletwo
\def\@oddhead{\hbox{} 
\rightmark\hfil\ninerm\thepage}
\def\@oddfoot{}\def\@evenhead{\ninerm\thepage\hfil 
\leftmark\hbox{}}\def\@evenfoot{}
\def\sectionmark##1{}\def\subsectionmark##1{}}

\begin{document}

\newcommand{\symbolfootnote}{\renewcommand{\thefootnote}
	{\fnsymbol{footnote}}}

\renewcommand{\thefootnote}{\fnsymbol{footnote}}
\newcommand{\alphfootnote}
	{\setcounter{footnote}{0}
	 \renewcommand{\thefootnote}{\sevenrm\alph{footnote}}}

\newcounter{sectionc}\newcounter{subsectionc}\newcounter{subsubsectionc}
\renewcommand{\section}[1] {\vspace{0.6cm}\addtocounter{sectionc}{1}
\setcounter{subsectionc}{0}\setcounter{subsubsectionc}{0}\noindent
	{\bf\thesectionc. #1}\par\vspace{0.4cm}}
\renewcommand{\subsection}[1] {\vspace{0.6cm}\addtocounter{subsectionc}{1}
	\setcounter{subsubsectionc}{0}\noindent
	{\it\thesectionc.\thesubsectionc. #1}\par\vspace{0.4cm}}
\renewcommand{\subsubsection}[1]
{\vspace{0.6cm}\addtocounter{subsubsectionc}{1}
	\noindent {\rm\thesectionc.\thesubsectionc.\thesubsubsectionc.
	#1}\par\vspace{0.4cm}}
\newcommand{\nonumsection}[1] {\vspace{0.6cm}\noindent{\bf #1}
	\par\vspace{0.4cm}}

\newcounter{appendixc}
\newcounter{subappendixc}[appendixc]
\newcounter{subsubappendixc}[subappendixc]
\renewcommand{\thesubappendixc}{\Alph{appendixc}.\arabic{subappendixc}}
\renewcommand{\thesubsubappendixc}
	{\Alph{appendixc}.\arabic{subappendixc}.\arabic{subsubappendixc}}

\renewcommand{\appendix}[1] {\vspace{0.6cm}
        \refstepcounter{appendixc}
        \setcounter{figure}{0}
        \setcounter{table}{0}
        \setcounter{equation}{0}
        \renewcommand{\thefigure}{\Alph{appendixc}.\arabic{figure}}
        \renewcommand{\thetable}{\Alph{appendixc}.\arabic{table}}
        \renewcommand{\theappendixc}{\Alph{appendixc}}
        \renewcommand{\theequation}{\Alph{appendixc}.\arabic{equation}}
        \noindent{\bf Appendix \theappendixc #1}\par\vspace{0.4cm}}
\newcommand{\subappendix}[1] {\vspace{0.6cm}
        \refstepcounter{subappendixc}
        \noindent{\bf Appendix \thesubappendixc. #1}\par\vspace{0.4cm}}
\newcommand{\subsubappendix}[1] {\vspace{0.6cm}
        \refstepcounter{subsubappendixc}
        \noindent{\it Appendix \thesubsubappendixc. #1}
	\par\vspace{0.4cm}}

\def\abstracts#1{{
	\centering{\begin{minipage}{30pc}\tenrm\baselineskip=12pt\noindent
	\centerline{\tenrm ABSTRACT}\vspace{0.3cm}
	\parindent=0pt #1
	\end{minipage} }\par}}

\newcommand{\bibit}{\it}
\newcommand{\bibbf}{\bf}
\renewenvironment{thebibliography}[1]
	{\begin{list}{\arabic{enumi}.}
	{\usecounter{enumi}\setlength{\parsep}{0pt}
\setlength{\leftmargin 1.25cm}{\rightmargin 0pt}
	 \setlength{\itemsep}{0pt} \settowidth
	{\labelwidth}{#1.}\sloppy}}{\end{list}}

\topsep=0in\parsep=0in\itemsep=0in
\parindent=1.5pc

\newcounter{itemlistc}
\newcounter{romanlistc}
\newcounter{alphlistc}
\newcounter{arabiclistc}
\newenvironment{itemlist}
    	{\setcounter{itemlistc}{0}
	 \begin{list}{$\bullet$}
	{\usecounter{itemlistc}
	 \setlength{\parsep}{0pt}
	 \setlength{\itemsep}{0pt}}}{\end{list}}

\newenvironment{romanlist}
	{\setcounter{romanlistc}{0}
	 \begin{list}{$($\roman{romanlistc}$)$}
	{\usecounter{romanlistc}
	 \setlength{\parsep}{0pt}
	 \setlength{\itemsep}{0pt}}}{\end{list}}

\newenvironment{alphlist}
	{\setcounter{alphlistc}{0}
	 \begin{list}{$($\alph{alphlistc}$)$}
	{\usecounter{alphlistc}
	 \setlength{\parsep}{0pt}
	 \setlength{\itemsep}{0pt}}}{\end{list}}

\newenvironment{arabiclist}
	{\setcounter{arabiclistc}{0}
	 \begin{list}{\arabic{arabiclistc}}
	{\usecounter{arabiclistc}
	 \setlength{\parsep}{0pt}
	 \setlength{\itemsep}{0pt}}}{\end{list}}

\newcommand{\fcaption}[1]{
        \refstepcounter{figure}
        \setbox\@tempboxa = \hbox{\tenrm Fig.~\thefigure. #1}
        \ifdim \wd\@tempboxa > 6in
           {\begin{center}
        \parbox{6in}{\tenrm\baselineskip=12pt Fig.~\thefigure. #1 }
            \end{center}}
        \else
             {\begin{center}
             {\tenrm Fig.~\thefigure. #1}
              \end{center}}
        \fi}

\newcommand{\tcaption}[1]{
        \refstepcounter{table}
        \setbox\@tempboxa = \hbox{\tenrm Table~\thetable. #1}
        \ifdim \wd\@tempboxa > 6in
           {\begin{center}
        \parbox{6in}{\tenrm\baselineskip=12pt Table~\thetable. #1 }
            \end{center}}
        \else
             {\begin{center}
             {\tenrm Table~\thetable. #1}
              \end{center}}
        \fi}

\def\@citex[#1]#2{\if@filesw\immediate\write\@auxout
	{\string\citation{#2}}\fi
\def\@citea{}\@cite{\@for\@citeb:=#2\do
	{\@citea\def\@citea{,}\@ifundefined
	{b@\@citeb}{{\bf ?}\@warning
	{Citation `\@citeb' on page \thepage \space undefined}}
	{\csname b@\@citeb\endcsname}}}{#1}}

\newif\if@cghi
\def\cite{\@cghitrue\@ifnextchar [{\@tempswatrue
	\@citex}{\@tempswafalse\@citex[]}}
\def\citelow{\@cghifalse\@ifnextchar [{\@tempswatrue
	\@citex}{\@tempswafalse\@citex[]}}
\def\@cite#1#2{{$\null^{#1}$\if@tempswa\typeout
	{IJCGA warning: optional citation argument
	ignored: `#2'} \fi}}
\newcommand{\citeup}{\cite}

\def\fnm#1{$^{\mbox{\scriptsize #1}}$}
\def\fnt#1#2{\footnotetext{\kern-.3em
	{$^{\mbox{\sevenrm #1}}$}{#2}}}

\font\twelvebf=cmbx10 scaled\magstep 1
\font\twelverm=cmr10 scaled\magstep 1
\font\twelveit=cmti10 scaled\magstep 1
\font\elevenbfit=cmbxti10 scaled\magstephalf
\font\elevenbf=cmbx10 scaled\magstephalf
\font\elevenrm=cmr10 scaled\magstephalf
\font\elevenit=cmti10 scaled\magstephalf
\font\bfit=cmbxti10
\font\tenbf=cmbx10
\font\tenrm=cmr10
\font\tenit=cmti10
\font\ninebf=cmbx9
\font\ninerm=cmr9
\font\nineit=cmti9
\font\eightbf=cmbx8
\font\eightrm=cmr8
\font\eightit=cmti8


\newcommand\ba{\begin{array}}
\newcommand\ea{\end{array}}
\newcommand\ben{\begin{equation}}
\newcommand\een{\end{equation}}
\newcommand\beq{\begin{equation}}
\newcommand\eeq{\end{equation}}
\newcommand\bea{\begin{eqnarray}}
\newcommand\eea{\end{eqnarray}}

\newcommand{\sinc}{{\rm sinc}}


\begin{flushright}
Imperial/TP/94-95/47\\
{\tt hep-ph/9510385}\\
20th October 1995\\
\end{flushright}
\centerline{\tenbf VORTEX DENSITIES AND CORRELATIONS AT PHASE TRANSITIONS
\footnote{Invited talk at the Fourth Thermal Fields Workshop, Dalian
(China), August 1995. E-mail; R.Rivers@IC.AC.UK}
}
\baselineskip=16pt
\vspace{0.8cm}
\centerline{\tenrm R.J. RIVERS}
\baselineskip=13pt
\centerline{\tenit Blackett Lab., Imperial College, Prince Consort Road}
\baselineskip=12pt
\centerline{\tenit London, SW7 2BZ, U.K.}

\vspace{0.9cm}
\abstracts{We present a model for the formation of relativistic
vortices (strings)
at a quench, and calculate their density and correlations. The significance
of these to early universe and condensed-matter
physics is discussed.
}

\vfil
\twelverm   
\baselineskip=14pt
\vspace*{-0.7cm}
\vspace*{-0.35cm}
\vglue 0.3cm
\vglue 0.4cm
\vglue 1pt


\section{\bf Introduction}

Despite our improved understanding of the very early universe the
nature of the phase transitions in the Grand Unification era is
largely unknown.  This would seem to be a serious drawback in
our modelling of the universe
since, in all but the baldest inflationary scenarios, an imprint of
these transitions is visible today.
In particular, it has been argued by Kibble\cite{kibble1} and
others\cite{shellard}
that the large-scale structure of
the universe can be attributed to cosmic strings (vortices in the
fields) formed at that time.

However, it has been suggested that the initial conditions of any
string network are largely washed out after a few expansion times,
at which the network is assumed to approach a scaling regime with a
few large loops and long strings per horizon volume continuing to produce
smaller
loops by self and mutual intersection.  This has seemed to obviate
the need for any detailed description of the early microscopic
dynamics that set the boundary conditions for the latter classical
picture.

Despite this, there are two circumstances in which this lack of
detailed initial information leaves us at a loss.  The first
concerns the {\it density} of the defects formed.  Although not of
immediate importance to the cosmologists, for the reasons given
above, it does relate directly to the
interesting proposition\cite{zurek1} that the
production
of vortices in superfluid $^{4}He$  and $^{3}He$
may share many attributes with the production of cosmic strings.  To
date, the primary data from such experiments concern defect
densities\cite{lancaster,helsinki}.  If there is a similarity
between the production mechanisms in these different regimes of high
$T$ (low chemical potential) and low $T$ (high chemical potential)
we should be able to predict the density of
vortices in both superfluid experiments and the early universe.

Secondly, and of direct importance to the astrophysicists, the scaling
solutions
for the early universe mentioned above require the presence of some
{\it 'infinite'} string, {\it i.e.}, vortices that do not self-intersect.  The
presence of such
string is, in part, determined by the initial conditions and it is
important to know whether it is present for reasonable models.  For
example, it has been suggested that vortices produced by bubble
nucleation in a strong first-order transition will {\it only} form small
loops\cite{borrill}.
The tendency for string to form loops should be visible in string
density correlation functions.  In a condensed matter context, the
same correlation functions should enable us to
estimate the superflow that would occur at a superfluid
quench from fluctuations alone\cite{zurek1}.

In this talk I shall show how these problems can be addressed in a model
of vortex formation by unstable long-wavelength  Gaussian
fluctuations at a temperature 'quench'.  For simplicity, flat spacetime is
assumed.
Relying on a slow 'rollover', this excursionary model can only describe weak
coupling
systems for the short times while the domains are growing
before the defects freeze out.  Although this is
unsatisfactory for most early universe applications and for
low-temperature many-body systems, we know in principle\cite{boyanovsky}
how to include back-reaction (still within the
context of a Gaussian approximation) to slow down domain growth
prior to
the field fluctuations spreading to the ground-state manifold. If
the results of Ref.7 are a reliable guide to our model then the
conclusions that we shall draw are likely to survive.
Greater detail is given in recent work by
myself and my collaborators,
Tim Evans, Alasdair Gill and Glykeria Karra\cite{alray,alray2},
from which this talk is drawn.

\section{\bf Vortex Distributions}

Consider the theory
of a complex scalar field $\phi ({\bf x},t)$.  The complex order
parameter of the theory is $\langle\phi\rangle = \eta e^{i\alpha}$ and the
theory possesses
a global $O(2)$ symmetry that we take to be broken
at its (continuous) phase transition.

Initially, we take the system to be in the symmetry-unbroken
(disordered) phase, in which the field is distributed about $\phi = 0$
with zero mean.
We assume that, at some time $t = t_0$,
the $O(2)$ symmetry of the ground-state (vacuum) is broken by a rapid change in
the
environment inducing an explicit time-dependence in the field
parameters. Once this quench is completed the $\phi$-field potential
$V(\phi ) = -M^{2} |\phi |^{2} + \lambda|\phi |^{4}$ is
taken to have the familiar symmetry-broken form with $M^{2}> 0$.

In practice we expect that, as the complex scalar field begins to
fall from the false ground-state into the true ground-state,
different points on the ground-state manifold (the circle $S^{1}$,
labelled by the phase $\alpha$ of $\langle\phi\rangle$) will be chosen at each
point in space.
If this is so then continuity and single valuedness will sometimes force the
field to
remain in the false ground-state at $\phi = 0$ . For example, the phase of
the field may change by an integer multiple of $2 \pi$ on going round
a loop in space. This requires at least one
{\it zero} of the field within the loop, each of which has topological
stability
and characterises a vortex (or string).  As to the density of the
strings, if the phase $\alpha$ is correlated over a distance $\xi$,
then the density of strings passing through any surface will be
$O(\xi^{-2})$ {\it i.e.} a fraction of a string per unit correlation area.
On the completion of the
transition a network of strings survives whose further evolution is
determined by classical considerations as the  field gradients
adjust to minimise the energy.

The question is, how can we infer these late-time
string densities and the density correlations from the microscopic field
dynamics?
The answer lies in the fact,
noted earlier, that the
string core is a line of zeroes of the fields $\Phi_{a}$ ($a$=1,2).
This is equally true for both relativistic and non-relativistic
$O(2)$ theories.  The problem is solved if we can identify those zeroes which
will freeze
out to define the late-time vortices.
This will require careful winnowing, since it is apparent
 that quantum fluctuations lead to zeroes
of the fields on all distance scales (even in the disordered phase).
However, most of these zeroes will be transient.  Two levels of
screening are required before the relevant zeroes can be identified.
Firstly, those zeroes whose positions vary rapidly, or which
annihilate one another on time scales
of $O(m^{-1})$, where $m$ provides the mass-scale,
can be discounted.  Secondly, since $m^{-1}$ also measures the
typical vortex width, zeroes on scales smaller than this should also
be ignored.
For the moment we ignore this problem, and count {\it every} zero.

To see how to proceed,
consider an ensemble of systems evolving from one of a set of disordered
states whose relative probabilities are known,
to an ordered state as indicated above.
As a prologue to the problem in hand, counting the zero {\it line}
densities appropriate to vortices, we
consider the much simpler case of counting the zeroes of a real field $\Phi
(x)$
in {\it one}
\footnote{Because of the peculiarities of one spatial dimension,
that would confuse the issue, we pretend that we are examining a
one-dimensional subset of a real field in higher dimensions and
ignoring other degrees of freedom.}
space dimension.
  On implementing the change of state on the line, these zeroes
fluctuate and annihilate, but some of them will come to define the
position of
'kinks' (at which $\Phi '(x) >0$), some the position of 'antikinks'
(at which $\Phi '(x) <0$), the one-dimensional counterparts of
vortices and 'anti'-vortices.

Suppose, at a given time, the zeroes of $\Phi (x)$ occur at $x = x_1
,x_2, ...$.
It is useful to define {\it two} densities.
The first,
\ben
{\bar\rho}(x) = \sum_{i} \delta (x - x_{i}),
\een
is the {\it total} density of zeroes, not distinguishing between kinks and
antikinks (by which we now mean zeroes at which the field has
positive or negative derivative). The second is the {\it topological} density,
\ben
\rho (x) = \sum_{i} n_{i}\delta (x - x_{i}),
\een
where $n_{i} = sign(\Phi '(x_{i}))$, measuring (net) topological
charge, the number of kinks minus the number of antikinks.

Equivalently, in terms of the $\Phi$-field, the total density is
\ben
{\bar\rho}(x) = \delta [\Phi (x)]|\Phi '(x)|,
\label{rhob}
\een
since $\Phi '(x)$ is the Jacobian of the transformation from zeroes
to fields.  Similarly, the topological density is
\ben
\rho (x) = \delta [\Phi (x)]\Phi '(x).
\een

Analytically,
it is not possible to keep track of individual transitions, but we
can construct ensemble averages.
If the phase change begins
at time $t_{0}$ then, for $t > t_{0}$, it is  possible in principle
to calculate the probability $p_{t}[\Phi]$ that
$\phi (x, t)$ (the counterpart of $\phi ({\bf x},t)$)
takes the value $\Phi (x)$ at time $t$.
Ensemble averaging $\langle F[\Phi ]\rangle_{t}$ at time $t$ is understood
as averaging over the field probabilities
$p_{t}[\Phi ]$.
This is not
thermal averaging since we are out of equilibrium.

The situation we have in mind is one in which, for early times when
the available space  permits many domains,
\ben
\langle\rho (x)\rangle_{t} = 0.
\een
{\it i.e.} an equal likelihood of a kink or an antikink occuring in
an infinitesimal length, compatible with an initially
disordered state.  However, the density
correlation functions
\ben
C(x ;t) = \langle\rho (x)\rho (0)\rangle_{t}
\label{c1}
\een
will be non-zero, as will
\ben
{\bar n}(t)= \langle {\bar \rho} (x)\rangle_{t}.
\label{n1}
\een

Let us now turn to the complex field $\phi ({\bf x}, t)$ and its
more complicated vortices.
As before we assume that it is  possible in principle
to calculate the probability $p_{t}[\Phi]$ that
$\phi ({\bf x}, t)$ takes the value $\Phi ({\bf x})$ at time $t$
\footnote{Throughout, it will be convenient to decompose $\Phi$ into
real and imaginary parts as $\Phi  = \frac{1}{\sqrt{2}}(\Phi_{1} +
i\Phi_{2})$ (and $\phi$ accordingly).  This is because we wish to
track the field as it falls from the unstable ground-state hump at the centre
of the potential to the ground-state manifold in
Cartesian field space.}.
To generalise our observations about zeroes on the line to vortices, we
follow Halperin\cite{halperin} in defining the {\it
topological line density} ${\vec{\rho}}(\bf r)$ by
\ben
{\vec{\rho}}({\bf r}) = \sum_{n}\int ds \frac{d{\bf R}_{n}}{ds}
\delta^{3} [{\bf r} - {\bf R}_{n}(s)].
\label{corrr}
\een
In (2.1) $ds$ is the incremental length along the line of zeroes ${\bf
R}_{n}(s)$ ($n$=1,2,.. .) and $\frac{d{\bf R}_{n}}{ds}$ is a unit
vector pointing in the direction which corresponds to positive
winding number.

It follows that, in terms of the zeroes of $\Phi ({\bf r})$,
$\rho_{i}({\bf r})$ can be written as
\ben
\rho_{i}({\bf r}) = \delta^{2}[\Phi ({\bf r})]\epsilon_{ijk}\partial_{j}
\Phi_{1}({\bf r}) \partial_{k}\Phi_{2}({\bf r}),
\label{rho}
\een
where
$\delta^{2}[\Phi ({\bf r})] = \delta[\Phi_{1} ({\bf r})] \delta[\Phi_{2}
({\bf r})]$.
The coefficient of the $\delta$-function in Eq.\ref{rho} is
the Jacobian of the more complicated transformation from line zeroes to field
zeroes.
We shall also need the {\it total line density} $\bar{\rho}({\bf
r})$, the counterpart of $\bar{\rho}(x)$ of Eq.\ref{rhob},
\ben
\bar{\rho_{i}}({\bf r}) = \delta^{2}[\Phi ({\bf r})]|\epsilon_{ijk}\partial_{j}
\Phi_{1}({\bf r}) \partial_{k}\Phi_{2}({\bf r})|.
\label{rhobar}
\een

As before, ensemble averaging $\langle F[\Phi ]\rangle_{t}$ at time $t$
means averaging over the field probabilities
$p_{t}[\Phi ]$.
Again we assume
\ben
\langle\rho_{j}({\bf r})\rangle_{t} = 0.
\een
{\it i.e.} an equal likelihood of a string or an antistring passing
through an infinitesimal area.
However,
\ben
{\bar n}(t) = \; \langle\bar{\rho_{i}}({\bf r})\rangle_{t} \; > 0
\label{n3}
\een
and measures the {\it total} string density in the direction $i$, without
regard to string orientation.  The isotropy of the initial state
guarantees that ${\bar n}(t)$ is independent of the direction $i$.
Further, the line density
correlation functions
\ben
C_{ij}({\bf r} ;t) = \langle\rho_{i}({\bf r})\rho_{j}({\bf 0})\rangle_{t}
\een
will be non-zero, and give information on the persistence length of
strings.

In  practice, our ability to construct $p_{t}[\Phi ]$ over the whole
timescale $t > t_0$ from initial quantum fluctuations to late time
classicality is severely limited.  It is convenient to divide time
into four intervals, in each of which we adopt a different approach.
If  $m$ sets the mass-scale, there is an initial period
$t_0 < t < t_i = O(m^{-1})$, before which the field
is able to respond to the quench, however rapidly  it is
implemented, and which we can largely ignore.
Assuming weak coupling, of which more later, this is followed by an
interval $t_i < t < t_f$ in which, provided the quench is
sufficiently rapid, domains in field phase form,
grow, and then cease growing  as the field magnitude begins to
approach the $S^1$ groundstates.  Vortices will appear and be driven
apart by the coalescence of these domains.  In the third period the
field magnitude relaxes to the ground state values {\it i.e.} the vortices
freeze in. Finally, in the
last period, the vortices behave semiclassically.
A phenomenological description of this period based on
time-dependent Landau-Ginzburg theory has been given by Liu and
Mazenko\cite{maz}, in which they fill in and extend the work of
Halperin, and from whose calculations we borrow. See also Bray\cite{bray}.

Our immediate aim is limited to discussing the second period $t_i < t <
t_f$ on the (as yet unproven) assumption that the distribution of
relevant zeroes  at time $t_f$ is left largely unchanged by their freezing in.
This distribution of vortices can then be taken as initial data for
the final evolution of the network.  Although we have long term
plans for a semi-analytical linkage of this final stage to the
initial microscopic dynamics, at the
very least we can compare the data to that used in numerical
simulations.

\section{\bf A Gaussian Model for Vortex Distribution}

We have yet to specify the nature of the quench but it is already
apparent that, if we are to make further progress, additional approximations
are necessary.
We return to our one-dimensional example.
The most important concerns the independence, or approximate
independence, of the fields and their derivatives at the same point.
If that holds then ${\bar n}(t)$ of Eq.\ref{n1} separates as
\ben
{\bar n}(t)\approx\langle\delta [\Phi (x)]\rangle_{t} \langle |\Phi '(x)|
\rangle_{t}.
\label{n11}
\een
and we can estimate, or bound, each factor separately.  In fact, we
can perform explicit calculations if $p_{t}[\Phi]$ is {\it Gaussian}, for
which the approximate equality in Eq.\ref{n11} becomes exact, and this we shall
assume henceforth.  That this is not a frivolous exercise in solving
what we can solve but a representation of reasonable dynamics will be
shown, in part, later, for the 'slow-roll' dynamics that we
shall adopt.

Specifically, suppose that (for the one-dimensional case)
$\Phi$ is a Gaussian field for which
\ben
\langle\Phi (x)\rangle_{t} = 0 =
\langle\Phi (x)\Phi '(x)\rangle_{t},
\label{g11}
\een
and that
\ben
\langle\Phi (x)\Phi (y)\rangle_{t} = W (|x - y|;t).
\label{g22}
\een
All other connected correlation functions are taken to be zero.
Then all ensemble averages are given in terms of $W(r;t)$ which,
from the closed timepath integral formalism, can
be equally understood as the equal-time Wightman function,
\ben
W (|x - y|;t) = \langle\phi (x, t)\phi (y, t)\rangle
\label{w1}
\een
with
the given initial conditions.  In our case, where we shall assume thermal
equilibrium initially, this is the usual thermal Wightman function.
It is straightforward to see that
\ben
{\bar n}(t) =
\frac{1}{\pi}\biggl |\frac{W''(0;t)}{W(0;t)}\biggr |^{\frac{1}{2}}.
\label{nii}
\een
where primes on $W$ denote differentiation with respect to $x$.
The numerator in Eq.\ref{nii} is the second factor in Eq.\ref{n1}.
The  denominator is the $\delta$-function term, as follows directly
on writing $\langle\delta [\Phi (x)]\rangle_{t}$ as
$\int d\alpha\langle e^{i\alpha\Phi (x)}\rangle_{t}$.  On using the
same exponentiation it takes only a little manipulation to cast
the correlation function $C(x;t)$ of Eq.\ref{c1} in the form
\ben
C(x;t) = \frac{\partial h(x; t)}{\partial x},
\een
where
\ben
h(x;t) = \frac{-W'(x;t)}{2\pi\sqrt{W(0;t)^{2} - W(x;t)^{2}}}.
\label{h}
\een

The extension to $O(2)$ line zeroes is messy, but leads to no surprises.
Specifically, suppose that
\ben
\langle\Phi_{a}({\bf r})\rangle_{t} = 0 =
\langle\Phi_{a}({\bf r})\partial_{j}\Phi_{b}({\bf r})\rangle_{t},
\label{g1}
\een
and, further, that
\ben
\langle\Phi_{a}({\bf r})\Phi_{b}({\bf r}')\rangle_{t} = W_{ab}(|{\bf r} -{\bf
r} '|;t)
= \delta_{ab} W(|{\bf r} -{\bf r} '|;t),
\label{g2}
\een
is diagonal.  As before, all other connected correlation functions are taken to
be zero.

The density calculation proceeds as before.
It follows\cite{halperin,maz} that
\ben
{\bar n}(t) =
\frac{1}{2\pi}\biggl |\frac{W''(0;t)}{W(0;t)}\biggr |,
\label{ni}
\een
The line density-line density correlation
functions
have a much more complicated realisation\cite{maz}, albeit still in terms of
$h(r;t)$ of Eq.\ref{h}, but we shall not consider them here in any detail.

The folly of counting {\it all} zeroes is now apparent in the
ultraviolet divergence of $W(r;t)$ at $r=0$ in all dimensions.
None of the expressions given above is well-defined.  To identify
which zeroes will turn into our vortex network requires
coarse-graining, determined by the dynamics.

\section{Gaussian Dynamics and its Coarsening}

It is not difficult to justify our adoption of Gaussian field
fluctuations for the early period of vortex production.
We have already assumed that the initial conditions correspond
to a disordered state.  In the absence of any compelling evidence to
the contrary we achieve this by adopting thermal equilibrium at
a temperature $T$ higher than the critical temperature $T_{c}$
for $t < t_{0}$.  The potential
describing this state behaves, near the origin in field space, as
\ben
V(\phi ) = \frac{1}{2}m^{2}(T)\phi_{a}^{2}
\label{vg}
\een
for an effective mass $m(T)$, with $m^{2}(T)>0$ .
For the sake of argument we take $V(\phi )$ to be {\it exactly} as in
Eq.\ref{vg}.  The resulting Gaussian field distribution will not be
seriously disturbed by  weak coupling.  Further, as we shall see later,
initial conditions generally give slowly varying behaviour in the
correlation function, in contrast to the rapid variation due to the
subsequent dynamics, and calculations are insensitive to them.

In order to have as simple a change of phase as possible, we
assume an idealised quench, described by
giving a time-dependence to the effective mass $m(T)\equiv m(t)$
so that, at
$t = t_{0}$, $m^{2}(t)$ changes sign everywhere. Given that the
resulting theory displays symmetry-breaking, this change in sign in
$m^{2}(t)$ can be interpreted as due to a reduction in temperature.
Even more, we assume that, for $t > t_{0}$, $m^{2}(t)$
takes the {\it negative} value $m^{2}(t) = - M^2 <0$ {\it
immediately}, where $ - M^2$ is the mass parameter of the (cold)
relativistic Lagrangian.  That is,
the potential at the origin has been instantaneously inverted,
breaking the global $O(2)$ symmetry, interpreted as a temperature
quench from high-T to, effectively, zero temperature.   If, as we shall
further assume,  the $\lambda |\phi|^{4}$ field coupling
is very weak then,
for times $M(t -t_{0}) <
\ln(1/\lambda)$, the $\phi$-field, falling down the hill away from
the metastable vacuum, will not yet have experienced the upturn of
the potential, before the point of inflection.
Thus, for these small time intervals,  $p_{t}[\Phi ]$ is Gaussian, as required.
Henceforth, we take $t_{0} = 0$.

For such a weakly coupled field
the onset of the phase transition at time $t=0$ is characterised by
the instabilities of long wavelength fluctuations permitting the growth of
correlations. Although the initial
value of $\langle \phi \rangle$ over any volume is zero, we
anticipate that
the resulting evolution will lead to
domains of constant $\langle \phi \rangle$ phase, whose boundaries will
trap vortices.

Consider small amplitude fluctuations of $\phi_a$, at the top of the
parabolic potential hill.  Long wavelength fluctuations, for which $|{\bf
k}|^2 < M^2$, begin to grow exponentially. If their growth rate
$\Omega (k) = \sqrt{M^2 - |{\bf k}|^2}$ is much slower than
the rate of change of the environment which is causing the quench,
then those long wavelength modes are unable to track the quench.
Unsurprisingly, the time-scale at which domains appear in this
instantaneous quench is $t_i = O(M^{-1})$. As long as the time taken
to implement the quench is comparable to $t_i$ and much less than
$t_f = O(M^{-1}ln (1/\lambda ))$ the approximation is relevant.

This picture essentially tells us how to coarsegrain the ultraviolet zeroes to
identify the vortices that will form the network.  Firstly, we note that if
$\Phi$ is Gaussian,
then so is the  coarsegrained field on scale $L$,
\ben
\Phi_{L}({\bf r}) = \int d^{3}r'\,I(|{\bf r} - {\bf r}'|)\Phi ({\bf r}'),
\een
where $I(r)$ is an indicator (window) function, normalised to unity,
which falls off rapidly for $r>L$.
The only change is that Eq.\ref{g2} is now replaced by
\ben
\langle\Phi_{L,a}({\bf r})\Phi_{L,b}({\bf r}')\rangle_{t} = W_{L,ab}(|{\bf r}
-{\bf r} '|;t)
= \delta_{ab} W_{L}(|{\bf r} -{\bf r} '|;t),
\label{gn2}
\een
where $W_{L}=\int\int IWI$ is now cut off at distance scale $L$.
$W_{L}(0;t)$, its derivatives, and all relevant quantities constructed
from $W_{L}$  are ultraviolet {\it finite}.  The distribution of
zeroes, or line zeroes, of $\Phi_L$ is given in terms of $W_{L}$ as
in the previous section.

Choosing $L = M^{-1}$ solves all our problems simultaneously.
At wavelengths $k^{-1} < L$ {\it i.e.} $k > M$) the field fluctuations
are oscillatory, with time scales $O(M^{-1})$.  Only those
long wavelength fluctuations with $k^{-1} > L$ have the steady exponential
growth
that can lead to the field migrating on larger scales to its
groundstates.  Further, as the field settles to its groundstates,
the typical vortex thickness is $O(M^{-1})$, and we only wish to
attribute {\it one} zero to each vortex.  By taking $L = M^{-1}$ we
are choosing not to count zeroes within a string, apart from the
central core.

We are now in a position to evaluate $p_t[\Phi]$, or rather $W_{L}(r;t)$,
which we now write as $W_{M}(r;t)$ for $t > 0$, and
calculate the defect density accordingly.
This situation of inverted harmonic oscillators was studied many
years ago by Guth and Pi\cite{guth} and Weinberg and Wu\cite{weinberg}.
In the context of domain formation, we refer to
the recent work of Boyanovsky et al.\cite{boyanovsky}, and our own\cite{alray}.
For weak coupling we recover what would have been our first naive guess
for the coarse-grained correlation
function $\langle\phi_{L,a}({\bf r},t)\phi_{L,b}({\bf
r}',t)\rangle$
based on the growth of the unstable modes $\phi_{a}({\bf
k},t)\simeq e^{\Omega (k)t}$ alone,
\ben
W_{M}(r;t)\simeq\int_{|{\bf k}|<M} d \! \! \! / ^3 k
\, e^{i {\bf k} . {\bf r} }
\;e^{2\Omega (k)t},
\label{Wapp}
\een
provided we have not quenched from too close to the transition
\footnote{In which case initial fluctuations are so large that the
field magnitude has already sampled the ground states and a slow
roll is inappropriate.}
and
$t>M^{-1}$ (preferably $t\gg M^{-1}$).
The reason why $W_{M}(r;t)$ is approximately independent of the initial
conditions
is that
the integral at time t is dominated by a
peak in the integrand $k^{2} e^{2\Omega (k)t}$ at $k$ around $k_c$, where
$t k_c^2\approx M$.
Any temperature dependence only occurs in
slowly varying factors (like $coth\beta\omega /2$) and can be ignored.

\section{Densities and Correlations}

In the first instance we understand the dominance of wavevectors at $k_{c}^{2}
= M/t$ in
the integrand as defining a length scale
\ben
\xi (t) = O(\sqrt{t/M}),
\label{xit}
\een
 once $Mt > 1$, over which the independently varying fields $\phi_a$
are correlated in magnitude.
However, in our introductory comments we anticipated that string
distributions would be determined by the length over which the {\it
phase} of the field is correlated.  For this simple model, with only
a single length, $\xi (t)$ has to serve both purposes.

To see how
this happens, we take
$\xi (t) = 2 \sqrt{t/M}$.
To calculate the number density of vortices at early times we
insert the expression Eq.\ref{Wapp} for $W_M$ into the equations derived
earlier, to find
\ben
{\bar n}(t) = \frac{1}{\pi}
\frac{1}{\xi(t)^2},
\label{nt}
\een
permitting us, from our introductory comments, to interpret $\xi (t)$ as a
correlation length for
phases.
We note that the dependence on time
$t$ of both the density and density correlations
 is only through the correlation length $\xi (t)$.
We have a {\it
scaling} solution in which, as the
domains of coherent field form and expand, the interstring distance
grows accordingly.  Since the only way the defect density can
decrease without the background space-time expanding is by
defect-antidefect annihilation, we deduce that the coalescence of
domains proceeds by the annihilation of small loops of
string, preserving roughly
one string zero per coherence area,  a long held belief for whatever
mechanism.

As well as determining the string density, we
are also looking for signs of anticorrelation, which enables us
to determine the persistence length of strings in the network ({\it i.e.}
how bendy they are).  The bendier, the more string that will occur
in small loops.  This is important, since the conventional string
model for large scale structure
formation in the universe requires a certain amount of infinite
string.
In fact, there may be no need to calculate the line density correlation
functions $C_{ij}$ to appreciate that there is a higher fraction of
string in loops than would have been anticipated from the early
simulations which laid out a regular domain structure\cite{tanmay}.
On inspection, the {\it broad} peaking in wavelength $l
= k^{-1}$ about $k=k_c$ in the integrand of Eq.\ref{Wapp} is
understood as indicating that the domains, both in field magnitude and
phase, of characteristic
linear dimension $\xi (t)$, have a substantial variation $\Delta\xi (t)$ in
size.
Simple calculation shows that
\ben
\frac{\Delta\xi (t)}{\xi (t)} = \frac{\Delta k}{k_{c}} = \frac{1}{2}.
\label{rms}
\een
independent of time.

To make use of Eq.\ref{rms} requires two further assumptions.  The
first, which can be checked using the methods of Ref.7, is that the
domains growing as $t^{\frac{1}{2}}$ stop expanding almost
immediately as soon as the field {\it magnitudes} probe the spinodal region and
the
instability switches off.
The second is that, during this time and the time it takes for the field
magnitude
to freeze on the $S^{1}$ groundstates,
$\Delta\xi /\xi$ of Eq.\ref{rms} describes (albeit approximately) the
variation in domain size over which field {\it phase} is correlated.
In
numerical simulation of string networks
the inclusion of variance in the size of field phase domains
 shows\cite{andy} that, the greater the
variance, the more string is in small loops.  The variance of
Eq.\ref{rms} suggests much more string ({\it e.g.} twice or more) in small
loops.
However, it is not easy to marry the somewhat different
distributions of this simulation to ours.  We can say no more than there
seems to be some infinite string and, in an early universe context,
some is enough.

To do a little better we need to evaluate density correlation functions.
For $W_M$ above we find that, in units of ${\bar n}^{2}(t)$, the
anticorrelation is large.
In particular, a calculation of $C_{33}$ for
separation $r$ in the 1-2 plane gives
\bea
C_{33}(r;t)&=& \frac{1}{r}\frac{\partial}{\partial r}h^{2}(r; t)
\nonumber
\\
&=& {\bar n}^{2}(t)\bigg[ -1 + O\bigg(\frac{r^{2}}{\xi^{2}(t)}\bigg)\bigg]
\label{c33}
\eea
when $r < \xi(t)$.
Although it is difficult to be precise, this again suggests a significant
amount of string in small loops.
For $r\gg\xi$, $C_{33}$ falls off
exponentially as
\ben
C_{33}(r;t) = O(e^{-2r^{2}/\xi^{2}(t)}).
\label{expc}
\een
showing that $\xi$ indeed sets the scale at which strings see one
another.
Unfortunately, it has not yet proved possible to turn expressions for the
$C_{ij}$ into statements about self-avoidance, fractal dimension, or
whatever is required to understand the length distribution of the
resulting string network.

However, we can use the correlation functions to determine
the variance in vortex winding number
$N_S$ through an open surface $S$ in the 1-2 plane, which we take
to be a disc of radius $R$.  There is a complication in that, as can
be seen from Eq.\ref{corrr},
$\rho_{3}$ counts zeroes weighted by the cosine of the angle with
which they pierce $S$, whereas winding number counts zeroes in $S$
without weighting.  The problem is therefore essentially a
two-dimensional problem.  Using topological charge conservation
across the whole plane, it is not difficult to see that
\ben
(\Delta N)_{S}^{2} = O({\bar n}^{2}\xi^{3}(t)R)= O(R/\xi (t)).
\een
All that is required is short range zero-density correlation functions.
Since $(\Delta N)_S$ is $(\Delta\alpha)_{S}/2\pi$, where
$(\Delta\alpha)_{S}$ is the variance in the field phase around the
perimeter $\partial S$ of $S$, this means in turn that
\ben
(\Delta\alpha)_{S}^{2} = O(R/\xi (t)).
\label{dal}
\een
This is naturally interpreted as a
random walk in {\it phase} along the perimeter with average step length
$\xi (t)$, within which the phase is correlated.
 Calculations will be given elsewhere\cite{alray2}.

Since only the boundary region of $S$  gives a
contribution to the variance, we would get the
same result if we were to quench in a annulus.  It has already been
proposed\cite{zurek1} that quenches in an annulus be performed
for superfluid $^{4}He$, for which $(\Delta\alpha)_{S}$ can
be identified with a supercurrent.

\section{Acknowledgements}

As well as thanking my collaborators mentioned earlier I would like to thank
Dan
Boyanovsky and Rich Holman for fruitful discussions since the meeting.
As a consequence this article has had some of the open ends of the original
talk tidied.
I would also like to thank the organisers, especially
Prof. Gui, for their hospitality and an enjoyable meeting.  Finally,
I would like to thank the Royal Society in particular for their
financial support.

\end{document}